\magnification=\magstep1
\hfuzz=6pt
\baselineskip=16pt

$ $
\vskip 1in
\centerline{\bf Decoherent histories and generalized measurements}

\vskip 1cm

\centerline{Seth Lloyd}

\centerline{MIT, Mechanical Engineering}

\vskip 1cm

\noindent{\it Abstract:} The theory of decoherent histories allows
one to talk of the behavior of quantum systems in the absence
of measurement.  This paper generalizes the idea of decoherent
histories to arbitrary open system operations and proposes
experimentally testable criteria for decoherence.

\vskip 1cm

The conventional answer to the question, `When is a quantum
system in a particular state?' is `When it has been measured
to be so.'  Situations frequently arise, however, in which 
no measurement apparatus is available, but we would still
like to talk about the behavior of quantum systems and
assign probabilities to different alternatives. 
One of the most successful methods for assigning probabilities
to different alternatives is supplied by the theory of decoherent 
histories [1-15].  This method, developed by Griffiths [1-2], Omn\`es [3-4], and
by Gell-Mann and Hartle [5-7], provides straightforward
mathematical criteria for when classical probabilities can be
assigned to quantum systems in the absence of measurement.
When a set of histories of a quantum system is decoherent, then one
can `talk about them at the dinner table,' even when no measurement
has taken place [1-2].  The theory of decoherent histories
is crucial for analysing quantum mechanical systems, such as the universe
taken as a whole, on which no measurements are made.  Originally formulated for
projections, the theory of decoherent histories was generalized by 
Kent [8-9] and by Rudolph [10-11] to apply more generally to positive operator 
valued measures (POVMs).
This paper extends the theory of decoherent histories to generalized
open system operations, and supplies physical and experimental criteria 
for when it is possible to talk about properties of 
quantum systems at the dinner table.    

The theory proposed here defines decoherence in terms of 
measurement.  If the act of measuring does not affect the
probabilities of subsequent measurements, then the histories
corresponding to that sequence of measurements will be
said to be decoherent.  For example, when we observe the flight
of a football through the air, the fact that our eyes measure
its trajectory has no effect on its future flight: the different
possible trajectories of the football are decoherent.
By contrast, in the double slit experiment, making a 
measurement of which slit the particle goes through
destroys the interference pattern; the histories corresponding 
the particle going through one slit or the other are coherent. 

A set of histories is decoherent if the probabilities
assigned to later alternatives do not depend on whether 
measurements corresponding to earlier alternatives are
made or not.  
As will be seen below, this measurement-based definition of
decoherence is equivalent to the standard definition in the
cases investigated heretofore.  In addition, this definition
allows the notion of decoherent histories to be extended
to sequences of generalized measurements. 

\bigskip\noindent{\it Quantum measurements:}
To make sense of the proposed measurement-based criterion, 
review the effect of measurement on quantum systems [12].
First, take the case of the conventional Copenhagen approach to
measurement.  Consider a
quantum system whose dynamics is given by a unitary  $U$ on a Hilbert 
space ${\cal H}$ that takes density matrices $\rho \rightarrow U 
\rho U^{\dagger}$.  This is the deterministic time evolution
desribed by the Schr\"odinger equation.  (The same formalism allows one to 
treat nonunitary, trace-preserving, completely positive,
`super-scattering' operators corresponding to open quantum
systems, by the well-known method of embedding this evolution
in a unitary interaction between the system and a quantum
environment [12]).  A projective or von Neumann measurement
on the system corresponds to a set of projection operators $\{P_i \}$.
The projection operators are Hermitian $P_i = P_i^{\dagger}$,
mutually exclusive, $P_iP_j= \delta_{ij} P_i$, and
exhaustive, $\sum_i P_i = 1$.   The measurement gives the result
$i$ with probability $p_i={\rm tr} P_i \rho$, in which case
the system is in the state $\rho_i=(1/p_i) P_i \rho P_i$. 

The Copenhagen interpretation of quantum
mechanics has the considerable advantage of predicting accurately
the probabilistic behavior of quantum systems when measured.  But it has the 
disadvantage of introducing two types of time evolution, one
for quantum systems on their own, and a second for quantum systems
interacting with a measuring apparatus.
In contrast to the Copenhagen interpretation, which assigns
probabilities only through measurement,
the decoherent histories approach to
quantum systems specifies necessary and sufficient conditions
for assigning probabilities to events in the history of
a given quantum system in terms of the system's dynamics
alone, without recourse to interactions
between the system and a classical measuring device.
The basic idea of the decoherent histories approach
is to identify criteria for when it is possible to 
assign probabilities to events in time that obey the
normal laws of probability, so that probabilities
sum to one, and the probability $p(i)$ for an alternative $i$ 
can be expressed in terms of probability
sum rules for alternatives at different times: $p(i)=\sum_j p(ij)$,
where $j$ labels set of alternatives at a different time.

The decoherent histories approach revolves around the
decoherence functional for sequences of events.  Just
as in the picture of quantum measurement given above,
in the decoherent histories approach, an alternative at time $t_j$ corresponds
to a projection operator $P^j_{i_j}$, where $\{P^j_{i_j} \}$ is
an exhaustive, mutually exclusive set of projection operators
corresponding to the different alternatives that could occur
at time $t_j$.  Define the decoherence functional
$$D(i_1\ldots i_n; i'_1\ldots i'_n) = {\rm tr}
P^n_{i_n}U \ldots  U P^1_{i_1}\rho
P^1_{i'_1} U^\dagger \ldots U^\dagger P^n_{i'_n}\quad. \eqno(1)$$
\noindent The on-diagonal part of the decoherence functional,
$D(i_1\ldots i_n; i_1\ldots i_n)$, is equal to the probability 
$p(i_1\ldots i_n)$
that one would obtain the result $i_1\ldots i_n$
if one were to make measurements corresponding
to the sets of projections  $\{P^j_{i_j} \}$
at times $t_j$.  The significance of the off-diagonal part
is as follows.

The general condition under which the
alternatives in a set of histories can be assigned probabilities can
readily be seen to be
$$ {\rm Re} D(i_1\ldots i_n; i'_1\ldots i'_n) \propto
\delta_{i_1i'_1}\ldots \delta_{i_ni'_n},\eqno(2)$$
\noindent for all $i_k, i'_k$, in which case the
histories are said to be (weakly) decoherent [1-7].
The reason for adopting this definition of decoherence
is physical.  In addition to having a mathematically well-defined
definition of decoherence, it is useful to have a definition
that can be tested experimentally.  The definition of
decoherence has an explicit experimental interpretation.  To check
for decoherence, prepare two sets of systems in the state $\rho$;
on one set, make projective measurements corresponding
to $P^j_{i_j}$, and in the other set do not.
Then make measurements corresponding to the
$P^\ell_{i_\ell}$ on both sets and compare whether the results of
this measurement have the same probabilities on both sets.  For example,
in the double slit experiment, one cannot assign probabilities
to the particle going through one slit or the other, because
a projective measurement that determines which slit it goes through
also destroys the interference pattern generated by the slits:
the double slit experiment is intrinsically coherent.
In contrast, a sequence of positions of a particle that interacts strongly
with a bath of oscillators can be assigned probabilities as long
as the $P_j$'s corresponding to the position measurements do not
pin down its position too closely and the $t_j$ are spaced sufficiently
far apart: the sequence of approximate positions of the particle
are decoherent [5-7].   Of course, the experimentally-based definition
proposed here cannot be tested for systems such as the universe
as a whole on which it is not possible (for us, at any rate)
to make a projective measurement.  Nonetheless, in such
situtations the mathematical criterion for decoherence is still
well-defined.

Let us now slightly rephrase the condition for decoherence.
One can assign probabilities to the
event corresponding to  $P^j_{i_j}$ if the probabilities
for the outcomes of subsequent measurements are unchanged
whether or not one makes a measurement corresponding
to $\{P^j_{i_j} \}$ at time $t_j$ or not.  In other words,
we are allowed to talk about a quantum system having a certain
property if we could have measured that property without having
an effect on later probabilities.  More precisely,
let $\hat D_{S}$ be the decoherence functional derived from
$D$ by omitting some set $S$ of projections, e.g., 
$$\eqalign{&\hat D_{S}(i_1\ldots i_{j-1} i_{j+1}\ldots i_n; 
i'_1\ldots i'_{j-1} i'_{j+1}\ldots i'_n)\cr &= {\rm tr}
P^n_{i_n}U \ldots U P_{j+1}U^2P_{j-1} U\ldots  U P^1_{i_1}\rho
P^1_{i'_1} U^\dagger \ldots U^\dagger P_{j-1} {U^\dagger}^2 
P_{j+1}U^\dagger \ldots U^\dagger P^n_{i'_n}\cr}$$ corresponds
to a decoherence functional derived from $D$ by omitting
the set $S=\{P^j_{i_j}\}$ consisting of the $j$'th projections.  
Let $\bar D_{S}$ be the decoherence
functional derived from $D$ by tracing over the projections
in $S$, e.g.,
$$\eqalign{& \bar D_{S}(i_1\ldots i_{j-1} i_{j+1}\ldots i_n;
i'_1\ldots i'_{j-1} i'_{j+1}\ldots i'_n)\cr &= 
\sum_{i_j} D(i_1\ldots i_{j-1} i_j i_{j+1}\ldots i_n;
i'_1\ldots i'_{j-1} i_j i'_{j+1}\ldots i'_n)\cr}$$
If the diagonal elements of $\hat D_{S}$
are equal to the corresponding diagonal elements of $\bar D_{S}$,
then making a non-demolition measurement corresponding to
the projections in $S$ does not affect the probabilities
for later events.  In this case, we will say that the
system is decoherent with respect to $S$.  If the system
is decoherent with respect to all $S$ then we will say
that it is decoherent. 
It can easily be seen that demanding the equality of the 
diagonal elements of $\hat D_{S}$ 
to the corresponding diagonal elements of $\bar D_{S}$,
for all $S$, is equivalent to the decoherence condition (1).

As an example,
consider a spin 1/2 particle with trivial dynamics, $U= {\bf I}$.  
$P^x_\pm = (1/2)({\bf I}\pm \sigma_x)$ are projection
operators on the states spin $x = \pm 1/2$, similarly
for $P^y_\pm$.  Let the initial state of the spin be 
spin $z$ up $\rho = {\bf 1 + \sigma_z}/2$ 
It is easy to verify that the decoherence
functional $D(i,j;i'j') = {\rm tr} 
P^x_jP^y_i \rho P^y_{i'}P^x_{j'}$ exhibits (weak)
decoherence: for example $D(+,+;-,+) = i/4$.  
Accordingly, the probabilities for the final measurement of $x$ are
independent of whether or not one makes the initial measurement
of $y$.

Now consider the case of open-system operations in general,
including the case of positive operator valued measures
(or POVMs, as above).  A general open-system operation [12] is 
defined by a set of operators $A_{\mu i}$ such that 
$\sum_{\mu i} A^\dagger_{\mu i} A_{\mu i} = {\bf 1}$.
A generalized measurement corresponding to the $A$'s gives the result
$\mu$ with probability $p_\mu = {\rm tr} \sum_i A_{\mu i}
\rho A^\dagger_{\mu i}$.  In order to apply the experimentally-based
definition of decoherence, we must also specify the state
in which the open-system operation leaves the system.  In
analog with the projective case above, we will take the
state of the system after the measurement to be 
$\rho_\mu= (1/p_\mu) \sum_i A_{\mu i} \rho A^\dagger_{\mu i}.$
The $A$'s can be thought of as (the square roots of) an
exhaustive but non-exclusive set of projections.  All 
Markovian open-system operations on a quantum system can be shown
to be described by such a set of $A_{\mu i}$ [12].
Open-system operations describe the general time evolution of 
quantum system interacting with an environment with which
the system is initially uncorrelated.  This
environment could consist of a thermal bath, or a measuring
device, or of a generic quantum system.

Generalized measurements represent the most general situation 
in which one couples a quantum system
to another quantum system, then makes a projective measurement
on the two systems.  Projective measurements are a special
case of generalized measurement.  Generalized measurements are exhaustive (they
cover all possibilities) but not exclusive (two possibilities
are not mutually exclusive).  Accordingly, Hartle [6] has suggested
that some but not all generalized measurements
be thought of as `fuzzy' projections.

A simple example of such a `fuzzy' projection is the following.
Consider a quantum system such as a spin 1/2 particle with two 
states, $|\uparrow\rangle$ and $|\downarrow\rangle$.  Let 
$A_0 = |\uparrow\rangle \langle \uparrow| + (1/\sqrt 2)
|\downarrow\rangle\langle\downarrow|$ and $A_1 = 
(1/\sqrt 2) |\downarrow\rangle\langle\downarrow|$.
Clearly, $A^\dagger_0 A_0 + A^\dagger_1 A_1 = {\bf 1}$.
Suppose that the system is in the initial state
$\rho = (1/2)(  |\uparrow\rangle \langle \uparrow| +
|\downarrow\rangle\langle\downarrow|$.  The generalized
measurement corresponding
the the $A$'s will give the result 0 with probability
3/4 and the result 1 with probability 1/4.  When the
result is 1, the system is in the state 
$|\downarrow\rangle \langle\downarrow|$;
while when the result is 0, the system
is in the state $(2/3)|\uparrow\rangle \langle \uparrow| +
(1/3) |\downarrow\rangle\langle\downarrow|$.
That is, when the result is 1, the system is in the state
$|\downarrow\rangle$ with certainty; while when the result
is 1, it is in the state $|\uparrow\rangle$ with probability
$2/3$ and in the state $|\downarrow\rangle$ with probability
$1/3$.  Such a generalized measurement
corresponds to a measurement that yields
uncertain information about the system.  It is straight-forward
to design a measurement protocol that performs the generalized 
measurement in
question: for example, a measuring apparatus that makes an 
imprecise measurement of spin that half the time erroneously returns the
result $0$ when it the system is actually in the state $|\downarrow\rangle$,
implements this generalized measurement.

\bigskip\noindent{\it Decoherent histories for generalized measurements:}
The decoherence functional for open-system operations
can be defined in analog to the conventional decoherence
functional as [10-11]:
$$D(\mu_1i_1\ldots \mu_ni_n;\mu'_1 i'_1\ldots \mu'_n i'_n) = 
{\rm tr} A^n_{\mu_ni_n}U \ldots  U A^1_{\mu_1i_1}\rho
{A^1}^\dagger_{\mu'_1i'_1} U^\dagger \ldots U^\dagger 
{A^n}^\dagger_{\mu'_ni'_n}\quad. \eqno(3)$$
Rudolph [10-11] has suggested that in the case of open-system
operations, decoherence be defined in analog to the case
of projective measurements to be the situation in which
the real part of the off-diagonal terms of the decoherence
functional vanish.  (Rudolph also sets criteria in terms
of D-posets for when sequences of POVMs can be considered a 
`legal' sequence.)  This criterion is satisfying from the
mathematical point of view, as it allows one consistently
to assign probabilities to histories in the abstract sense.
However, the criterion has no obvious physical interpretation.
In particular, the physically motivated definition of decoherence
given above, in which decoherence is defined to be a situation
in which earlier measurements do not affect the results of
later measurements, is lost.

More explicitly, when decoherence is referred to general
open-system operations rather than to projection operators,
the two definitions of decoherence (1) and (2) above are
no longer equivalent.  In particular, the probabilities
for later measurements may remain unaffected by earlier
measurements even when the off-diagonal parts of the decoherence
functional fail to vanish.  The reason lies in the non-exclusive
nature of the $A$'s. (Rudolph also notes that the non-exclusive
nature of the $A$'s typically causes the real parts of the
off-diagonal terms in the decoherence functional not to
vanish.)  As a simple example, consider the simple
spin system with $U={\bf 1}$, $\rho={\bf 1}/2$.  Let the first
set of operations in the decoherence functional
be $A_0$ and $A_1$ defined as above and let the second
set of operations be the trivial operation ${\bf 1}$.
corresponding to making no measurement at all.  It is
easy to see that the probability for the second measurement
is independent of whether the first is made or not (the
second measurement has only one result, which occurs
with probability 1).  It is equally easy to see that
the real part of the off-diagonal term in the decoherence
functional fails to vanish:
$$D(1,0;1,1)= {\rm tr} {\bf 1}  A_0 {\bf 1}/2 A^\dagger_1 {\bf 1}   
= 1/2. \eqno(4)$$

In fact, the physical and experimental significance of the vanishing of the 
off-diagonal parts of the decoherence functional is unclear.
Accordingly, we return to our earlier physical intuition
behind the definition of decoherence: we ask that earlier
measurements not change the probabilities for later ones.
In other words, defining $\hat D_{S}$ for general open-system
operations to be $D$ with some set $S$ of open-system operations
omitted, and $\bar D_{S}$ to be the trace of $D$ over the set
$S$ we demand that the on-diagonal terms of $ \hat D_{S}$
corresponding to the results of measurements $\mu$
be equal to the corresponding on-diagonal terms of $\bar D_{S}$,
in which case we say that the system is decoherent with
respect to $S$.  If the system is decoherent with respect
to all $S$, then we say that it is decoherent, as above.
More precisely, our decoherence condition is that 
$$ \sum_{i_1\ldots i_N} 
\hat D_{S}(\mu_1 i_1\ldots\mu_Ni_N;\mu_1 i_1\ldots\mu_N i_N)
= \sum_{i_1\ldots i_N} 
\bar D_{S}(\mu_1 i_1\ldots\mu_Ni_N;\mu_1 i_1\ldots\mu_N i_N),
\eqno(5)$$
for all $S$.  This mathematical criterion
is equivalent to the physical one of demanding that the
probabilities for later measurements are independent of
whether earlier measurements were made or not.

As with projective measurements, it is straightforward
to test whether or not a system is decoherent with respect
to a set of operations $S$.  To check
for decoherence, prepare two sets of systems in the state $\rho$.
Then make the measurements corresponding to the 
$A^\ell_{\mu\ell i_\ell}$
On one set, make measurements corresponding
to the effects in $S$, and on the other set omit these measurements.
Compare whether the results of the remaining experiments have the
same probalities in both sets.  If they do, then the system
is decoherent.  If they do not, then the system is coherent.

This experimentally-based definition of decoherence in equation
(5) is no longer equivalent to the condition that the off-diagonal
terms in the decoherence functional vanish.  The reason is
that unlike the projectors $P_i$, the open-system operations
$A_{\mu i}$ do not sum to ${\bf 1}$.  Accordingly, the definition
of decoherence for open-system operations proposed in (5) is
different from that proposed by Rudolph.  Rudolph was in fact
aware that it can be possible to define probabilities consistently
in situations where the real part of the off-diagonal terms of the
decoherence functional fail to vanish.  The decoherent
histories defined here exploit exactly these situations.
In this sense, the definition for decoherent histories over
open-system operations defined here is complimentary to
Rudolph's approach.

Of course, there are situations where the two different definitions
coincide, as in the case when the $A$'s are projectors.
In general, when the effect that corresponds to not
distinguishing between alternative $i$, corresponding to
effect $A_i$, and alternative $j$, corresponding to effect $j$,
is equal to $A_i+A_j$, as is the case with projectors,
then vanishing of the off-diagonal terms of the decoherence
functional implies that the system is decoherent with respect
to measurement and {\it vice versa}.  In general, however, this 
is not the case.

The experimentally-based version of decoherent histories 
defined above is closer to the version proposed by
Kent [9].  Kent restricts himself to situations in which the
$A$'s are Hermitian (call such Hermitian effects $B$'s), and demands that 
$${\rm tr} B^n\ldots B^1\rho B^1\ldots B^n 
= \sum_{i_1\in I_1\ldots i_n\in I_n}
{\rm tr} B_{i_n}^n\ldots B_{i_1}^1\rho B_{i_1}^1\ldots B_{i_n}^n.\eqno(6)$$
Here $I_j$ represents some subset of the $\{i_j\}$ at
each point, and $B^j= \big( \sum_{i\in I_j} (B^j_i)^2\big) ^{1/2}$.
Our experimentally-based condition proposed in (5) is both
more general and less restrictive than Kent's definition.
It is more general in the sense that it applies to non-Hermitian
$A$'s.  It is less restrictive in the sense that, unlike
the Kent definition, it only requires equality
respect to $I_j=\{ i_j \}$ and not to arbitrary subsets.
Another way of putting it is that the experimentally-based
definition does not require that one define the effect
corresponding to $A_j$ `or' $A_{j'}$, which Kent's definition
requires.  Nonetheless, the two definitions are similar in
spirit.  In particular, in the case that the $A$'s are
Hermitian, then Kent's definition implies the experimentally
based definition (5): histories that are decoherent a l\`a Kent
are decoherent a l\`a experiment.  The converse, however, is not
true: there may be histories that experiment declares to be 
decoherent that are not decoherent a l\`a Kent.   
In the special case that the $A$'s are Hermitian
and there are only two alternatives at each point in time,
then the definitions are equivalent.  In general, however,
these definitions are not equivalent. Henceforth we will
stick to the experimentally based definition.

\bigskip\noindent{\it Examples:}
Now consider the following examples of decoherent histories
and generalized measurements.  
First take the case of a continuous system whose Hilbert
space is spanned by the states $|x\rangle$ that are eigenstates
of the operator 
$ X= \int_{-\infty}^\infty x |x\rangle\langle x| dx$.
A useful generalized measurement for such a system 
is the set of Gaussian quasi-projections [12]:
$A_{\mu} = (1/(2\pi)^{1/4}\Delta^{1/2})
\int_{-\infty}^{\infty} e^{-(x-\mu)^2/4\Delta^2}
|x\rangle \langle x| dx $
where the normalization is chosen so that ${\rm tr}
A^\dagger_\mu A_\mu = 1$ and $\int_{-\infty}^\infty
A^\dagger_\mu A_\mu d\mu = I$ where 
$I= \int_{-\infty}^\infty |x\rangle \langle x| dx $
is the identity operator (here there is
no need for the auxiliary index $i$).
This generalized measurement is frequently found in nature:
it corresponds to the measurement made when a pointer variable
with Gaussian fluctuations is coupled to the system variable $x$.

If we write the system density matrix in the $x$ basis
as $\rho=\sum_{xx'} \alpha_{xx'} |x\rangle\langle x'|$, then the
generalized measurement corresponding to the $A_\mu$ determines the value
of $\bar x= {\rm tr} \rho X$ to an accuracy $\Delta$. 
In addition, the measurement has the effect of reducing the
off-diagonal terms
of $\rho$ by an factor $e^{-(x-x')^2/2\Delta^2}$, corresponding
to a perturbation of size $\epsilon \approx \Delta X^2/2\Delta^2$,
where $\Delta X = \sqrt{{\rm tr} \rho X^2 - {\bar x}^2}$.

Regardless of the system's dynamics, the results above
imply that the system exhibits histories that are
approximately decoherent 
as long as $\Delta X << \Delta$.  
For example, if the system has Hamiltonian $P^2/M$ where
$P=-i\partial/\partial x$, and the initial state of the system
is a Gaussian wave packet, the system will exhibit
decoherent histories with respect to repeated applications
of the POVM up until the point that the usual $\sqrt t$
spreading of the wave packet makes $\Delta X \approx \Delta$.
Note that this is true even in the absence of an external
environment to decohere the system.  (Such an environment
enhances decoherence by removing off-diagonal terms
in the system density matrix.)

It is interesting to compare this type of `automatic'
decoherence with the well-known example of `automatic'
decoherence in the case of projections.
For projections, one can always guarantee
decoherent histories by projecting onto the time-evolved
version of the initial state.  Such a set of decoherent
histories has probability 1 for one history and probability
0 for the rest.  In the case of the `automatic' decoherence
exhibited by Gaussian quasi-projections, many histories
exhibit non-zero probability.  Indeed, under the condition
$\Delta X << \Delta$ that guarantees decoherence, at any instant
in time, if one were to make a measurement one would find
a spread in values for $\mu$ of $\approx \Delta$.  
In this example, decoherence is guaranteed because the
generalized measurement
localizes the system only approximately.  The resulting
uncertainty in the outcome of the measurement corresponding
to the generalized measurement can be thought of as a kind of `measurement
error' corresponding to finite precision on the part of
the apparatus.

Another useful generalized measurement is the set of projections onto spin
directions for a spin-1/2 particle: $A_{\hat u} = (1/2)
(I+\sigma_u)$, where $\sigma_{\hat u}= u_x \sigma_x 
+u_y \sigma_y + u_z \sigma_z $ is the generalized
Pauli matrix associated with spin along the axis $\hat u$.
Here we see that even in the case of the trivial dynamics
the spin is not decoherent with respect to this generalized
measurement:
taking the initial state to be the state
$\rho = |\uparrow\rangle_z\langle\uparrow|$, we note that 
$${\rm tr} \int_{{\hat u}_1} A_{{\hat u}_2} A_{{\hat u}_1} \rho 
A^\dagger_{{\hat u}_1} A^\dagger_{{\hat u}_2} 
\neq {\rm tr}A_{{\hat u}_2}\rho A^\dagger_{{\hat u}_2}, \eqno(6)$$ 
so that $D_{\hat 1} \neq D_{\bar 1}$.  Generalized measurements onto spin
directions disrupt the result of later generalized measurements onto spin
directions.  Interestingly, this result holds even when
the initial state is arbitrarily close to the identity. 
For example,
$\rho = (1-\epsilon)I + \epsilon |\uparrow\rangle_z\langle\uparrow|$
also fails to decohere even for the trivial dynamics.

This result has implications for NMR quantum computing.
Quantum computers are examples of systems whose histories
are highly coherent: they obtain their speed-ups over classical
computers by arranging interference between different
computational histories.  It has been suggested
that because nuclear spins are not entangled at room temperature,
room-temperature NMR quantum computing is not `truly
quantum,' that is, they do not rely on quantum-mechanical
coherence to obtain their results [16-17].  Decoherent histories
provide a way of determining whether or not a quantum
system exhibits coherence: if there is a set of decoherent
histories for a room-temperature NMR quantum computer
that reproduces the results of the computation at each
step, then the computer could be modeled by a classical
Markovian process.  The obvious set of operations with
respect to which one might define decoherent histories
for a quantum computer are the projections onto the
logical states $|0\rangle, |1\rangle$ of the computer's
quantum bits.  For an NMR quantum computer, these states
are usually identified with the states $|\uparrow\rangle_z,
|\downarrow\rangle_z$ of the spins in the molecule that
is performing the computation.  Now the question can
be asked, are the histories of these quantum bits,
or `qubits' decoherent?  The answer to this question
can easily be verified to be No, even when the state
of the molecule is arbitrarily close to equilibrium,
so that its states are not entangled: a projective
measurement made on the logical state of the spins
during the course of the quantum computation typically
puts the spins in an entirely mixed state and destroys
the results of the computation.

Can one verify this coherence experimentally by the
criterion for decoherence defined here?  That is,
can one perform projective measurements on
the state of the spins and see whether these measurements
affect the results of the quantum computation?  At first, the
answer might seem to be No, as it is not currently
possible to measure the state of individual spins
in room-temperature NMR.  
However, our criterion for experimentally-verifiable
decoherence has been described above in a way that
makes it appear more restrictive than it is.  Above, histories were defined
to be decoherent if it was possible to make projective
measurements corresponding to earlier events without
affecting the probabilities of later events.  Note
that this criterion does not require one to know the
results of those measurements.  That is,
what one wants to do to verify decoherence is to make
a projective measurement, and ignore the results.
Although it is not currently possible to make projective
experiments on individual spins at room temperature,
it is in fact possible to perform an operation that is
equivalent to making a projective measurement and
ignoring the results.  This operation corresponds simply
to dephasing the spins about the $z$-axis, which can
easily be done using, e.g., gradient pulses.
That is, testing for decoherent histories is equivalent to
subjecting the system to environmentally-induced
decoherence [18], and seeing whether this environmentally-induced
decoherence affects the results of later measurements.

To test whether a system such as a room-temperature
NMR quantum computer exhibits decoherent histories,
one simply performs the computation twice, once
with dephasing the qubits and once without.  
If the computation works just as well
with dephasing, then the histories are decoherent.
If it doesn't work as well, then the histories are
coherent.  As noted above, all known algorithms
for quantum computation are coherent to a high
degree, whether the states of the quantum computer
are entangled or not.

Quantum computers are coherent with respect to
qubit histories.  However, one might also imagine
that there exist some other histories with respect
to which the quantum computers decohere.  An
example of such histories are those corresponding to
the spin-direction generalized measurement defined above.  However,
as shown above, this generalized measurement does not exhibit decoherent
histories even for trivial time evolutions for single
spins. 

\bigskip\noindent{\it Summary:} This paper presented a definition of
decoherent histories in terms of generalized measurements.
A history is decoherent if earlier measurements
do not change the results of later measurements.
This measurement-based criterion for decoherence allows
ready experimental tests of whether histories are decoherent
or not.   

\vfill
\noindent{\it Acknowledgements:} The author would like to thank
C. Caves, M. Gell-Mann, J. Hartle, and J. Halliwell for helpful
discussions, and the Newton Institute for its hospitality during
the writing of this work.  
This work was supported by ARDA, DARPA, NSF, Hewlett-Packard, and
CMI. 

\vfill\eject

\noindent{\bf References:}

\noindent[1] R. Griffiths, {\it J. Stat. Phys.} {\bf 36}, 219 (1984).

\noindent[2] R. Griffiths, {\it Phys. Rev. A} {\bf 54}, 2759 (1996).

\noindent[3] R. Omn\`es, {\it J. Stat. Phys.} {\bf 53}, 893 (1988).

\noindent[4] R. Omn\`es, {\it The Interpretation of Quantum 
Mechanics,} Princeton University Press, Princeton, 1994.

\noindent[5] M. Gell-Mann and J. Hartle, in {\it Complexity, Entropy,
and the Physics of Information,} SFI Studies in the Sciences of 
Complexity, Vol. VIII, W. Zurek, ed., Addison Wesley, Reading, 1990,
p. 425.

\noindent[6] J. Hartle, {\it Phys. Rev. D} {\bf 10}, 3173 (1991).

\noindent[7] M. Gell-Mann and J. Hartle, {\it Phys. Rev. D} {\bf 47},
3345 (1993).

\noindent[8] F. Dowker and A. Kent, {\it Phys. Rev. Lett.} {\bf 75},
3038 (1995).

\noindent[9] A. Kent, {\it Physica Scripta} {\bf T76}, 78 (1998).

\noindent[10] O. Rudolph, {\it Int. J. Theor. Phys.} {\bf 35},
1581 (1997)

\noindent[11] O. Rudolph, {\it J. Math. Phys.} {\bf 37}, 5368, (1996).

\noindent[12] A. Peres, {\it Quantum Theory:
Concepts and Methods}, Kluwer, Hingham, 1995. 

\bigskip\noindent[13] J.J. Halliwell, {\it Phys. Rev. D}
{\bf 58} 105015 (1998), quant-ph/9805062;
{\it Phys. Rev. D} {\bf 60} 105031 (1999), quant-ph/9902008;
{\it Phys. Rev. Lett.} {\bf 83} 2481 (1999), quant-ph/9905094;
`Decoherent Histories for Spacetime Domains,' in
{\it Time in Quantum Mechanics,} edited by J.G.Muga, R. Sala Mayato
and I.L.Egususquiza (Springer, Berlin, 2001), quant-ph/0101099.

\bigskip\noindent[14] J.B. Hartle,  {\it Phys. Scripta T}{\bf 76}, 67 (1998),
gr-qc/9712001.
 
\bigskip\noindent[15] T.A. Brun, J.B. Hartle, {\it Phys. Rev. E}{\bf 59},
6370-6380 (1999), quant-ph/9808024;  
{\it Phys. Rev. D}{\bf 60}, 123503 (1999), quant-ph/9905079.

\noindent[16] S. Braunstein {\it et al.}, {\it Phys. Rev. Lett.}
{\bf 83}, 1054 (1999).

\noindent[17] R. Schack, C.M. Caves, {\it Phys. Rev. A}{\bf 60}, 
4354-4362 (1999), quant-ph/9903101.

\noindent[18] W.H. Zurek, {\it Phys. Today} {\bf 44}, 36 (1991).

\vfill\eject\end